# Variational Interpolation Algorithm between Weak- and Strong-Coupling Expansions

H. Kleinert*
*Institut für Theoretische Physik,*
*Freie Universität Berlin*
*Arnimallee 14, D - 14195 Berlin*

## Abstract

For many physical quantities, theory supplies weak- and strong-coupling expansions of the types $\sum a_n \alpha^n$ and $\alpha^p \sum b_n (\alpha^{-2/q})^n$, respectively. Either or both of these may have a zero radius of convergence. We present a simple interpolation algorithm which rapidly converges for an increasing number of known expansion coefficients. The accuracy is illustrated by calculating the ground state energies of the anharmonic oscillator using only the leading large-order coefficient $b_0$ (apart from the trivial expansion coefficent $a_0 = 1/2$). The errors are less than 0.5% for all $g$. The algorithm is applied to find energy and mass of the Fröhlich-Feynman polaron. Our mass is quite different from Feynman's variational approach.

*email: kleinert@einstein.physik.fu-berlin.de; URL: http://www.physik.fu-berlin.de/~kleinert

1) Recently, the Feynman-Kleinert variational approximation to path integrals [1] has been extended to a systematic variational perturbation expansion [2]. This expansion converges uniformly and fast (for the anharmonic oscillator like $e^{-\mathrm{const}\times \mathrm{N}^{1/3}}$ in the order $N$ of the approximation [3]). Due to the uniformity of the convergence, it has given rise to an efficient method for extracting strong-coupling expansions from a weak-coupling expansions [4–6].

For many physical systems, there exists an independent knowledge of expansion coefficients for weak and strong couplings. Important examples are most lattice models of statistical mechanics. The purpose of this note is to propose a simple algorithm by which the variational perburbation expansion can be used to find a systematic convergent interpolation between weak- and the strong-coupling expansions.

The algorithm is completely general and holds for any physical system whose quantities possess expansions in some coupling constant $\alpha$ of type $\sum a_n \alpha^n$, for weak and of the type $\alpha^p \sum b_n (\alpha^{-2/q})^n$ for strong couplings, where either or both of these expansions may have a zero radius of convergence.

A typical example is the ground state energy of the anharmonic oscillator with $p = 1/3, q = 3$. We shall use this example to illustrated the power of the algorithm by calculating this the leading large-order coefficient $b_0$, apart from the trivial coefficent $a_0 = 1/2$. The errors are less than 0.5% for all $g$.

To make a prediction, we apply the algorithm to the Fröhlich-Feynman polaron [8], [9], where we know for the ground state energy the lowest three perturbation coefficients for weak couplings as well as the first two strong-coupling coefficients; for the polaron mass, only two strong-coupling coefficients have been calculated.

Our interpolation results can be compared with Feynman's famous variational solution with interesting discrepancies for the mass, calling for a calculation of higher perturbation or strong-coupling coefficients.

2) Following the method explained in [3], we rewrite the weak coupling expansion of order $N$,

$$E_N = \sum_{n=0}^{N} a_n \alpha^n, \tag{1}$$

as

$$E_N^{\omega} = \omega^p \sum_{n=0}^{N} a_n \left(\frac{\alpha}{\omega^q}\right)^n \tag{2}$$

where $\omega$ is an auxiliary parameter whose value is eventually set equal to 1 and $p, q$ are two parameters to be determined by general properties of the strong-coupling expansion. We now replace $\omega$ by the identical expression

$$\omega \to \sqrt{\Omega^2 + \omega^2 - \Omega^2} \tag{3}$$

and reexpand $E_N^{\mathrm{w}}$ in powers of $\lambda$ treating $\omega^2 - \Omega^2$ as a quantity of order $\alpha$. The reexpanded series is truncating after the order $n > N$.

The resulting expansion has the form

$$W_N(\alpha, \Omega) = \Omega^p \sum_{n=1}^{N} a_n f_n(\Omega) \left(\frac{\alpha}{\Omega^q}\right)^n \tag{4}$$

where

$$f_n(\Omega) = \sum_{j=0}^{N-n} \binom{(p-qn)/2}{j} (-)^j \left(1 - \frac{\omega^2}{\Omega^2}\right)^j . \tag{5}$$

Forming the first and second derivatives of $W_N(\alpha, \Omega)$ with respect to $\Omega$ we find the positions of the extrema and the turning points. The smallest among these is denoted by $\Omega_N$. The resulting $W_N(\alpha) \equiv W_N(\alpha, \Omega_N)$ constitutes the desired approximation to the energy.

It is easy to take this approximation to the strong-coupling limit $\alpha \to \infty$. For dimensional reasons, $\Omega_N$ increases with $\alpha$ like $\Omega_N \approx \alpha^{1/q} c_N$, so that

$$W_N(\alpha, \Omega_N) \approx \alpha^{p/q} c_N^p w_N^{(0)} \tag{6}$$

where

$$w_N^{(0)} = \sum_{n=0}^{N} a_n f_n(\infty) \left(\frac{1}{c_N^q}\right)^n . \tag{7}$$

The full strong-coupling expression is obtained by writing $W_N(\alpha,\Omega) = \Omega^p w_N(\hat\alpha, \omega^2/\Omega^2)$, with $\hat\alpha \equiv \alpha/\Omega^q$, and expanding $w_n$ in powers of $\omega^2/\Omega^2$. which behaves for $\alpha \to \infty$ like $(1/c^2)(\alpha/\omega^q)^{-2/q}$. The result is

$$W_N(\alpha) = \alpha^{p/q}\left[b_0(c) + b_1(c)\left(\frac{\alpha}{\omega^q}\right)^{-2/q} + b_2(c)\left(\frac{\alpha}{\omega^q}\right)^{-4/g} + \dots\right] \tag{8}$$

with

$$b_n(c) = \frac{1}{n!} w_N^{(n)}(\hat\alpha, 0)\hat\alpha^{(2n-p)/q}|_{\hat\alpha=1/c^q} \tag{9}$$

and the superscript $(n)$ denotes the $n$th derivative with respect to $\hat\omega^2$.

The parameters $p$ and $q$ in the expansion (2) are now determined to render the correct leading and the successive powers of $\alpha$ in the strong-coupling expansion (8).

The leading coefficient $c_N$ in the optimal frequency $\Omega_n$ is found by searching for the extrema of the leading coefficient $b_0(c)$ as a function of $c$ and choosing the smallest of them.

Explicitly

$$b_n(c) = \sum_{l=0}^{N} a_l \sum_{j=0}^{N-n} \binom{(p-lq)/2}{j}\binom{j}{n}(-1)^{j-n} c^{p-lq-2n}. \tag{10}$$

Next we have to correct for the fact that for large but finite $\alpha$ $\Omega$ has corrections to the behavior $\alpha^{1/q}c$. The coefficient $c$ will depend on $\alpha$ like

$$c(\alpha) = c + c_1\left(\frac{\alpha}{\omega^q}\right)^{-2/q} + c_2\left(\frac{\alpha}{\omega^q}\right)^{-4/g} + \dots, \tag{11}$$

requiring a reexpansion of $c$-dependent coefficients $b_n^c$ in (8). The expansion coefficients $\gamma_n$ are determined by extremizing $b_{2n}(c)$. The final result can again be written in the form (8) with $b_n^c$ replaced by $b_n$ which are determined by the equations shown in Table 1. The two leading coefficients receive no correction and are omitted.

It is now obvious that the knowledge of any strong-coupling coefficients $b_0, b_1$ can be exploited to determine approximately further coefficients $a_{N+1}, a_{N+2}, \dots$ and thus carry $W_N(\alpha)$ to higher orders. We merely have to solve Eq. (10) for as many $b_n$ as are available.

3) The weak-coupling expansion of the anharmonic oscillator looks like (1) with $\alpha = g/4$ (for a potential is $gx^4/4$). The lowest coefficient has is given by the ground state energy and has the value $a_0 = 1/2$.

The strong-coupling behavior is known from general scaling arguments to start out like $g^{1/3}$ followed by powers of $g^{-1/3}, g^{-1}, g^{-5/3}$. Inspection of (8) shows that this corresponds to $p = 1$ and $q = 3$. The leading coefficient is known extremely accurately [7,5], $b_0 = 0.667\,986\,259\,155\,777\,108\,270\,962\,016\,919\,860\,\ldots$ . This is now used to determine an approximate $a_1$ (forgetting that we know the exact value $a_1^{\rm ex} = 3/4$). The energy (4) reads for $N = 1$:

$$W_1(\alpha, \Omega) = \frac{\Omega}{4} + \frac{1}{4\Omega} + \frac{a_1}{4\Omega^2}. \tag{12}$$

Equation (10) yields, for $n = 0$:

$$b_0 = \frac{c}{4} + \frac{a_1}{c^2}. \tag{13}$$

Differentiating $b_0$ with respect to $c$ we find $c = c_1 \equiv 2a_1^{1/3}$. Inserting this into (13) fixes $a_1 = (4/3b_0)^3 = 0.773\,970\,\ldots$, quite close to the exact value. With our approximate $a_1$ we calculate $W_1(\alpha, \Omega)$ at its minimum, where

$$\Omega_1 = \begin{cases} \frac{2}{\sqrt{3}}\omega \cosh\left[\frac{1}{3}\text{acosh}(g/g^{(n)})\right] & \text{for} \quad g > g^{(n)}, \\ \frac{2}{\sqrt{3}}\omega \cos\left[\frac{1}{3}\arccos(g/g^{(n)})\right] & \phantom{\text{for}} \quad g < g^{(n)}, \end{cases} \tag{14}$$

with $g^{(0)} \equiv 2\omega^3/3\sqrt{3}$. The result is shown in Fig. (1). Since the difference with respect to the exact solution would be to small to be visible on a direct plot of the energy, we display the ratio with respect to the exact energy $W_1(\alpha)/E^{\rm ex}$. The accuracy is everyhwere better than 99.5 %. For comparison, we also display the much worse (although also quite good) variational perturbation result using the exact $a_1^{\rm ex} = 3/4$.

4) Let us now turn to the polaron model. The Hamiltonian operator reads

$$H = \frac{\mathbf{p}^2}{2m_b} + \sum_{\mathbf{k}} \hbar\omega_o a_{\mathbf{k}}^\dagger a_{\mathbf{k}} + \sum_{\mathbf{k}} \left(V_{\mathbf{k}} a_{\mathbf{k}} e^{i\mathbf{k}x} + \text{h.c.}\right) \tag{15}$$

where $m_b$ is the effective mass of the electron in the conduction band, $\mathbf{p}$ is the electron momentum, $\omega_{\rm O}$ is the frequency of optical phonons which are created and annihilated by $a_{\mathbf{k}}^{\dagger}$ and $a_{\mathbf{k}}$, and

$$V_{\mathbf{k}} = -i\frac{\hbar\omega}{|\mathbf{k}|}\left(\frac{4\pi\alpha}{V}\right)^{1/2}\left(\frac{\hbar}{2m_b\omega_{\rm O}}\right)^{1/4} \tag{16}$$

specifies the electron-phonon interaction in the volume $V$. The Fröhlich coupling constant

$$\alpha = \frac{e^2}{\hbar c}\sqrt{\frac{m_b c^2}{2\hbar\omega_{\rm O}}}\left(\frac{1}{\varepsilon_\infty} - \frac{1}{\varepsilon_0}\right) \tag{17}$$

involves the fundamental constants $e, c, \hbar$ and the electronic and static dielectric constants $\varepsilon_\infty$ and $\varepsilon_0$, respectively. This form of $V_{\mathbf{k}}$ assumes the size of the polaron to be large with respect to the lattice spacing. It further ignores spin and relativistic effects and the dispersion of the electron band.

In natural units with $\hbar = c = m_b = \omega_{\rm O} = 1$, the partition function of the polaron in thermal equilibrium at a fixed temperature $T$ is described by the path integral

$$Z(\beta) = \int \mathcal{D}x(\tau)\exp\left[-\frac{1}{2}\int_0^\beta d\tau\dot{\mathbf{x}}^2 + \frac{\alpha}{2^{3/2}}\int_0^\beta\int_0^\beta d\tau d\tau'\frac{e^{-|\tau-\tau'|}}{|\mathbf{x}(\tau)-\mathbf{x}(\tau')|}\right] \tag{18}$$

where $\beta = 1/T$ is the inverse temperature (at Boltzmann constant $k_{\rm B} = 1$). The weak-coupling expansion of the energy of the polaron is known up to the order $\alpha^3$ [10]:

$$E^{\rm w} = -\alpha - 0.0159196220\alpha^2 - 0.000806070048\alpha^3 - O(\alpha^4). \tag{19}$$

For strong couplings, the energy is [11]

$$E^{\rm s} = -0.108513\alpha^2 - 2.836 - O(\alpha^{-2}). \tag{20}$$

The polaron mass has the corresponding expansions [12], [11]:

$$m^{\rm w} = 1 + \frac{\alpha}{6} + 0.02362763\alpha^2 + O(\alpha^4) \tag{21}$$

$$m^{\rm s} = 0.0227019\alpha^4 + O(\alpha^2). \tag{22}$$

Feynman was the first to find a uniform all-coupling constant expressions from a variational approximation to the path integral (18):

$$E^{\rm F} = \operatorname*{Min}_{v,w} \frac{3}{4v}(v-w)^2 - \frac{\alpha}{\sqrt{\pi}} \int_0^\infty \frac{d\tau e^{-\tau}}{\{w^2\tau + [(v^2-w^2)(1-\epsilon^{-\varepsilon v})/v]\}^{1/2}} \tag{23}$$

and

$$m^F = 1 + \frac{1}{3\sqrt{\pi}} \alpha v^3 \int_0^\infty \frac{d\tau \tau^2 e^{-\tau}}{\{\omega^2\tau + [(v^2-\omega^2)(1-e^{-\tau v})/v\}^{1/2}}, \tag{24}$$

the latter being evaluated at the parameters $v(\alpha), w(\alpha)$ obtained in minimizing $E^{\rm F}$. For weak coupling Feynman's expressions are exact only to the order $\alpha$. They have the expansions [10]:

$$E^{F,{\rm w}} = -\alpha - 0.012345\alpha^2 - 6.43434\times 10^{-4}\alpha^3 - 4.643\times 10^{-5}\alpha^4 - 3.93\times 10^{-6}\alpha^5 - \ldots \tag{25}$$

$$m^{F,{\rm w}} = 1 + \frac{\alpha}{6} + 2.469136\times 10^{-2}\alpha^2 + 3.566719\times 10^{-3}\alpha^3 + 5.073952\times 10^{-4}\alpha^4 + \ldots \tag{26}$$

For strong couplings, the expansions are

$$E^{F,s} \approx -0.106103\alpha^2 - 2.8294 - 4.86399/\alpha^2 - 34.1952/\alpha^4 + \ldots \tag{27}$$

$$m^{F,s} \approx 0.020141\alpha^4 - 1.012775\alpha^2 + 11.85579 + \ldots \tag{28}$$

With the help of the interpolation algorithm based on the variational perturbation expansion we shall find new expressions for $E$ and $m$ which share with Feynman's the validity for all $\alpha$, but are more reliable at small and large $\alpha$ by possessing the presently most precise weak- and strong-coupling expansions (19), (20) and (21), (22).

5) We now apply our interpolation algorithm the expansions (19) and (20) for the energy. To make the series start out with $\alpha^0$ as required by the general ansatz (2), we remove an overall factor $-\alpha$ from $E$ and deal with $-E/\alpha$.

Then we see from (20) that the correct leading power in the strong-coupling expansion requires taking $p=1, q=1$. The knowledge of $b_0$ and $b_1$ allows us to extend the known weak coupling expansion (19) by two further expansion terms. Their coefficients $a_3, a_4$ are solutions of the equations

$$b_0 = \frac{35}{128} a_0 c + a_1 + \frac{15}{8}\frac{a_2}{c} + \frac{2a_3}{c^2} + \frac{a_4}{c^3} \tag{29}$$

$$b_1 = \frac{35}{32}\frac{a_0}{c} - \frac{5}{4}\frac{a_2}{c^3} + -\frac{a_3}{c^3}. \tag{30}$$

The constant $c$ governing the growth of $\Omega_N$ for $\alpha \to \infty$ is obtained by extremizing $b_0$ in $c$, which yields the equation

$$\frac{35}{128}a_0 - \frac{15}{8}\frac{a_2}{c^2} - \frac{4a_3}{c^3} - \frac{4a_4}{c^5} = 0. \tag{31}$$

The simultaneous solution of (29)—(31) renders

$$\begin{aligned} c_4 &= 0.09819868, \\ a_3 &= 6.43047343 \times 10^{-4}, \\ a_4 &= -8.4505836 \times 10^{-5}. \end{aligned} \tag{32}$$

The reexpanded energy (4) reads explicitly (for $E$ including the earlier-removed factor $-\alpha$)

$$\begin{aligned} W_4(\alpha,\Omega) = {}& a_0\alpha\left(-\frac{35}{128}\Omega - \frac{35}{32\Omega} + \frac{35}{64\Omega^3} - \frac{7}{32\Omega^5} + \frac{5}{128\Omega^7}\right) - a_1\alpha^2 \\ & + a_2\alpha^3\left(-\frac{15}{8\Omega} + \frac{5}{4\Omega^3} + \frac{-3}{8\Omega^5}\right) + a_3\alpha^4\left(-\frac{2}{\Omega^2} + \frac{1}{\Omega^4}\right) - a_4\alpha^5\frac{1}{\Omega^3} \end{aligned} \tag{33}$$

Extremizimg this we find $\Omega_4$ as a function of $\alpha$ [it turns out to be quite well approximated by the simple function $\Omega_4 \approx c_4\alpha + 1/(1+0.07\alpha)$]. This is to be compared with the optimal frequency obtained from minimizing the lower approximation $W_2(\alpha,\Omega)$:

$$\Omega_2^2 = 1 + \frac{4a_2}{3a_0}x^2 + \sqrt{\left(1 + \frac{4a_2}{3a_0}x^2\right)^2 - 1}, \tag{34}$$

which behaves likes $c_2\alpha + 1 + \dots$ with $c_2 = \sqrt{8a_2/3a_0} \approx 0.120154$. The resulting energy is shown in Fig. 2, where it is compared with the Feynman variational energy. For completeness, we have also plotted the weak-coupling expansion, the strond-coupling expansion, the lower approximation $W_2(\alpha)$ , and two Padé approximants which were given in the last of Refs. [10] as upper and lower bounds to the energy.

6) Consider now the polaron mass, where the strong-coupling behavior (22) fixes $p = 4, q = 1$. The coefficient $b_0$ allows us to determine an approximate coefficient $a_3$ and to calculate the variational perturbation expansion $W_3(\alpha)$. From (10) we find the equation

$$b_0 = -a_1c^3/8 + a_3c, \tag{35}$$

whose minimum lies at $c_3 = \sqrt{8a_2/3a_0}$ [this value follows, of course, also directly from (37], where $b_0 = \sqrt{32a_3^3/27a_1}$. Using $b_0$ from (22), we obtain $a_3 = [27a_1 b_0^2/32]^{1/3} \approx 0.0416929$.

From (4) we have

$$W_3(\alpha, \omega) = a_0 + a_1\alpha\left(-\frac{\Omega^3}{8} + \frac{3\Omega}{4} + \frac{3}{8\Omega}\right) + a_2\alpha^2 + a_3\alpha^3\Omega. \tag{36}$$

This is extremal at

$$\Omega_3^2 = 1 + \frac{4a_3}{3a_1}x^2 + \sqrt{\left(1 + \frac{4a_3}{3a_1}x^2\right)^2 - 1}. \tag{37}$$

The approximation $W_3(\alpha) = W_3(\alpha, \Omega_3)$ for the polaron mass is shown in Fig. 3, where it is compared with the weak and strong-coupling expansions and with Feynman's variational result. To see better the differences between the strongly rising curves, we have divided out the asymptotic behavior $M_{\rm as} = 1 + b_0\alpha^4$ before plotting the data. As for the energy, we have again displayed two Padé approximants given by the last of Refs. [10] as upper and lower bounds to the energy. Note that our interpolation differs considerably from Feynman's and higher order expansion coefficients in the weak or the strong coupling expansions will be necessary to find out which is the true behavior of the model.

Our curve has, incidentally, the strong-coupling expansion

$$m^{\rm s} = 0.0227019\alpha^4 + 0.125722\alpha^2 + 1.15304 + O(\alpha^{-2}), \tag{38}$$

the $\alpha^2$-term being in sharp contrast with Feynman's expression (28). On the weak-coupling side, a comparison of our expansion with Feynmans's in Eq. (26) shows that our coefficent $a_3 \approx 0.0416929$ is about 10 times larger than his.

Both differences are the reason for our curve forming a positive arch in Fig. 2, whereas Feynman's has a valley. It will be interesting to find out how the polaron mass really behaves. This would be possible by calculating a few more terms in either the weak- or the strong-coupling expansion.

Note that our interpolation algorithm is much more powerful than Padé's. First, we can account for an arbitrary fractional leading power behavior $\alpha^p$ as $\alpha \to \infty$. Second, the successive lower powers in the strong-coupling expansion can be spaced by an arbitrary $2/q$. Third, our functions have in general a cut in the complex $\alpha$-plane approximating the cuts in the function to be interpolated [13]. Padé approximants, in contrast, have always an integer power behavior in the strong-coupling limit, a unit spacing in the strong-coupling expansion, and poles to approximate cuts.

Acknowledgement

The author thanks Prof. J.T. Devreese for sending him a perprint of his new review article *Polarons*.

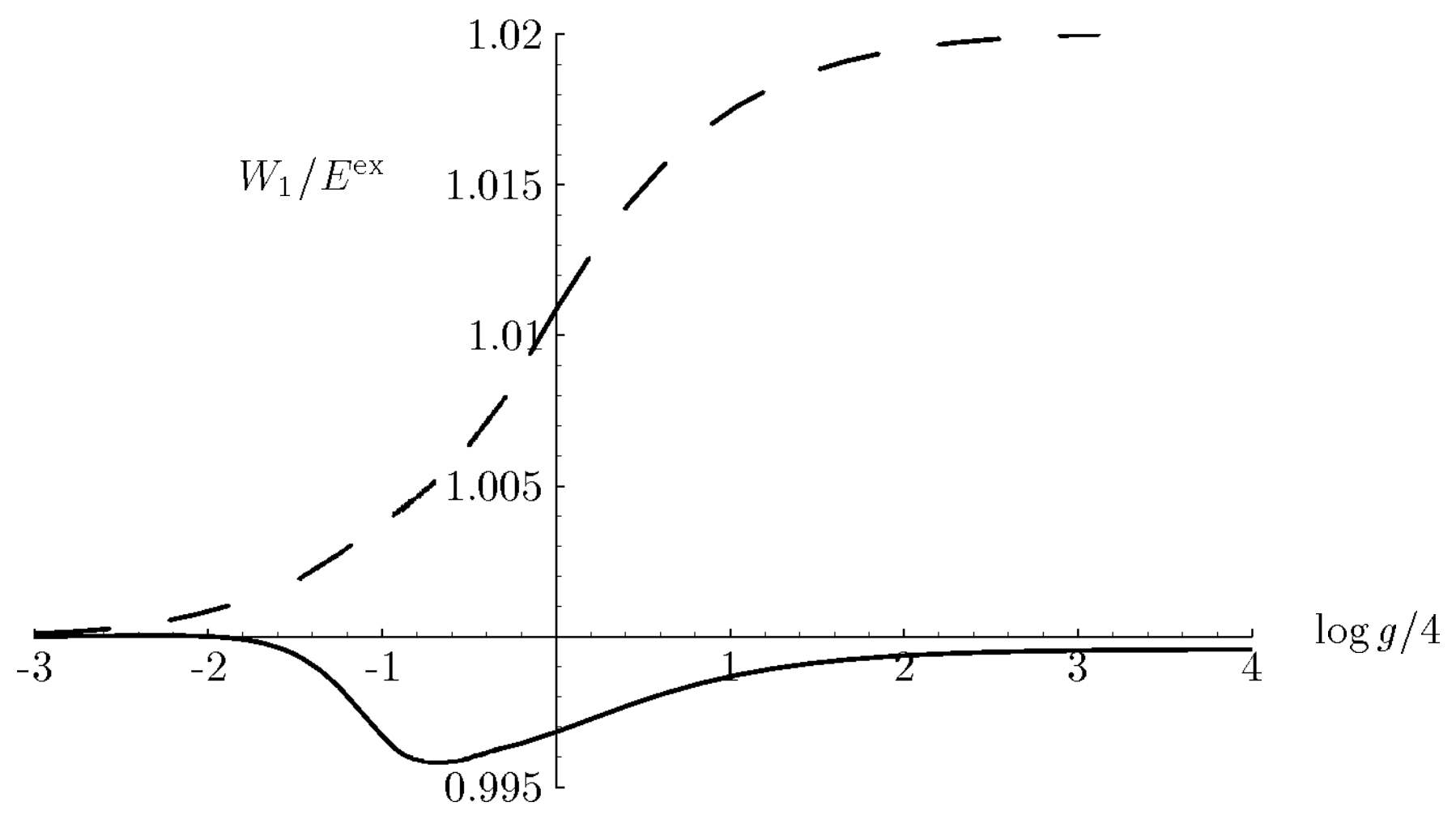

FIG. 1. Plot of the ratio of the intepolation energy with respect to the exact energy as a function of the coupling constant. The accuracy is everyhwere better than 99.5 %. For comparison, we also plot the variational perturbation result using the exact $a_1 = 3/4$.

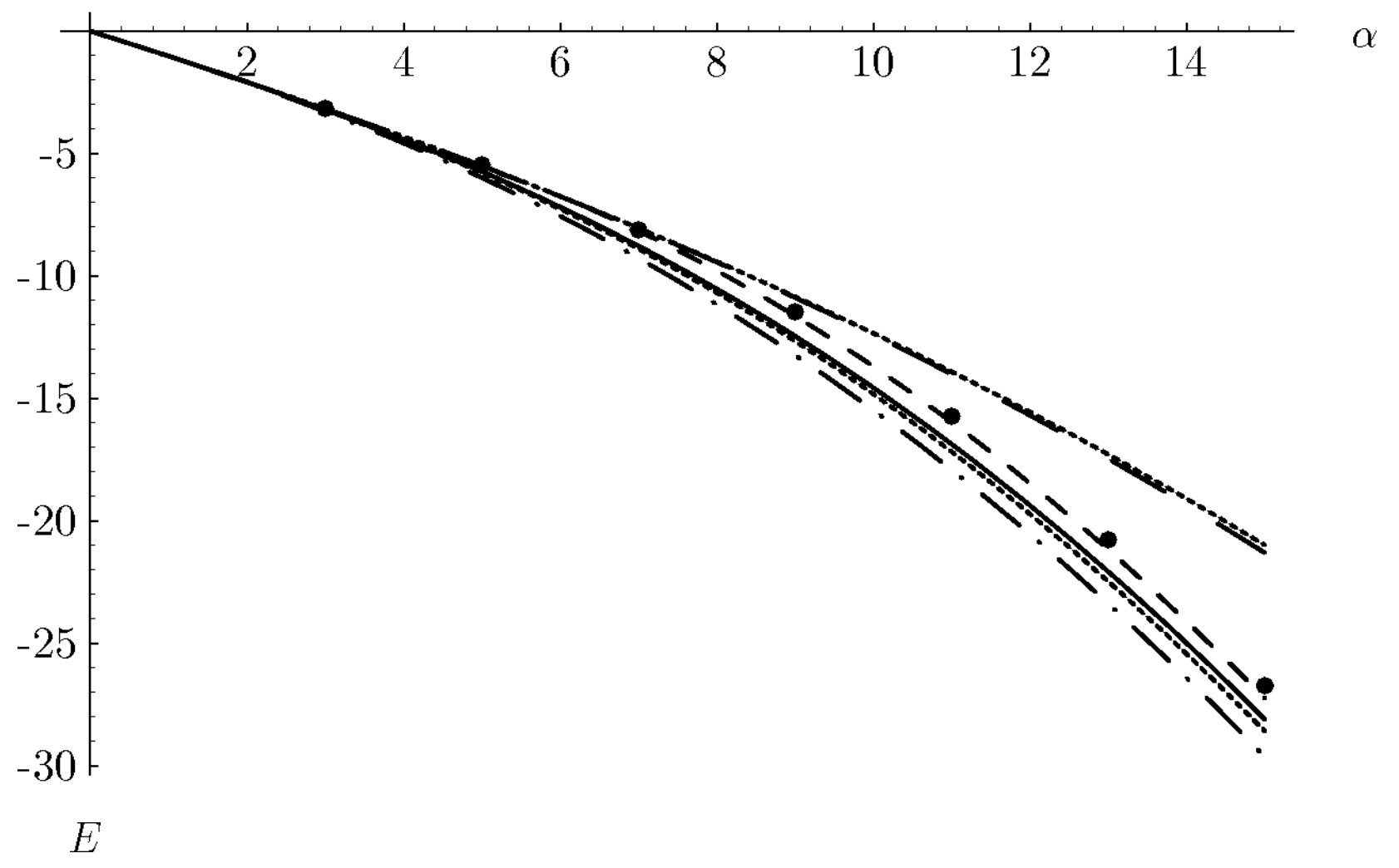

FIG. 2. The polaron energy obtained from our variational interpolation between the weak coupling expansion (dashed) and the strong-coupling expansion (short-dashed) to be compared with Feynman's variational approximation (fat dots). The dotted curves are Padé approximants which were given in the last of Refs. [10] as upper and lower bounds to the energy.

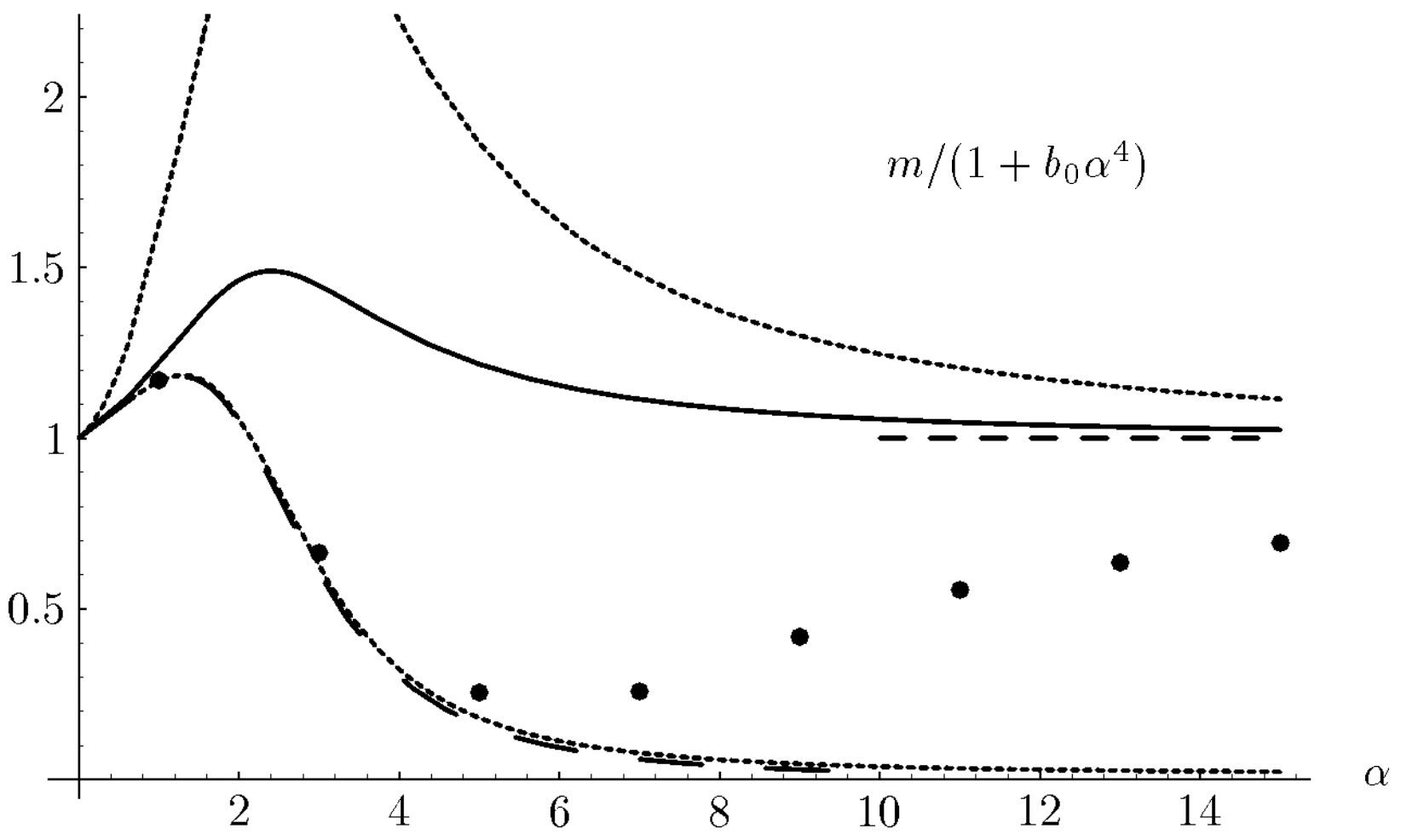

FIG. 3. The polaron mass curve interpolating optimally between the weak-(dashed) and strong-coupling expansions (short-dashed). To see better the differences between the strongly rising functions, we have divided out the asymptotic behavior $M_{\rm as} = 1 + b_0\alpha^4$ before plotting the curves. The fat dots show Feynman's variational approximation. The dotted curves are Padé approximants which were given in the last of Refs. [10] as upper and lower bounds to the mass.

# TABLES

TABLE I. Equations determining the coefficients $b_n$ in the strong-coupling expansion from the functions $b_n(c)$ and their derivatives. For brevity, we have suppressed the argument $c$ in the entries.

| $n$ | $b_n$ | $-c_{n-1}$ |
|---|---|---|
| 2 | $b_2 + c_1 b_1' + \frac{1}{2} c_1^2 b_0''$ | $b_1'/b_0''$ |
| 3 | $b_3 + c_2 b_1' + c_1 b_2' + c_1 c_2 b_0'' + \frac{1}{2} c_1^2 b_1'' + \frac{1}{6} c_1^3 b_0^{(3)}$ | $b_2' + c_1 b_1'' + \frac{1}{2} c_1^2 b_0^{(3)})/b_0''$ |
| 4 | $b_4 + c_3 b_1' + c_2 b_2' + c_1 b_3' + (\frac{1}{2} c_2^2 + c_1 c_3) b_0''$ | $(b_3' + c_2 b_1'' + c_1 b_2'' + c_1 c_2 b_0^{(3)}$ |
| | $+c_1 c_2 b_1'' + \frac{1}{2} c_1^2 b_2'' + \frac{1}{2} c_1^2 c_2 b_0^{(3)} + \frac{1}{6} c_1^3 b_1^{(3)} + \frac{1}{24} c_1^4 b_0^{(4)}$ | $+\frac{1}{2} c_1^2 b_1^{(3)} + \frac{1}{6} c_1^3 b_0^{(4)})/b_0''$ |